\documentstyle[epsf,prl,aps]{revtex}

\def\beq{\begin{equation}}
\def\eeq{\end{equation}}
\def\beqa{\begin{eqnarray}}
\def\eeqa{\end{eqnarray}}
\def\bI{\hbox{$\,I\!\!\!\!-$}}
\def\a{\alpha}
\def\b{\beta}
\def\p{\partial}
\def\e{\epsilon}

\def\r{\rho}
\def\vp{\varphi}
\def\O{\Omega}

\def\D{\Delta}
\def\th{\theta}
\def\Th{\Theta}
\def\t{\tilde}

\def\b{\bar}
\def\n{\nabla}
\def\ra{\rightarrow}

\def\four{\hbox{${}^{(4)}\!$}}
\def\five{\hbox{${}^{(5)}\!$}}
\def\six{\hbox{${}^{(6)}\!$}}
\def\bxi{\mbox{\boldmath $\xi$}}
\def\br{\mbox{\boldmath $r$}}

\def\bv{\mbox{\boldmath $v$}}
\def\bX{\mbox{\boldmath $X$}}

\begin{document}

\draft
\preprint{YITP--99--48}

\title{Almost analytic solutions to equilibrium sequences of irrotational
binary polytropic stars for n=1}

\author{Keisuke Taniguchi$\ddagger$ and Takashi Nakamura
}

\address{Yukawa Institute for Theoretical Physics,
	Kyoto University, Kyoto 606-8502, Japan}

\date{\today}

\wideabs{
\maketitle

\begin{abstract}

A solution to an equilibrium of irrotational binary polytropic stars
in Newtonian gravity is expanded in a power of $\e=a_0/R$, where $R$
and $a_0$ are the separation of the binary system and the radius of
each star for $R=\infty$. For the polytropic index $n=1$, the
solutions are given almost analytically up to order $\epsilon^6$. We
have found that in general an equilibrium solution should have the
velocity component along the orbital axis and that the central density
should decrease when $R$ decreases. Our almost analytic solutions can
be used to check the validity of numerical solutions.

\end{abstract}

}


Coalescing binary neutron stars (BNSs) are considered to be one of the
most promising sources of gravitational waves for laser
interferometers such as TAMA300, GEO600, VIRGO and
LIGO\cite{Thorne94}. We can determine the mass and the spin of neutron
stars from the gravitational wave signals in the inspiraling phase. We
may also extract the informations on the equation of state of a
neutron star from the signals in the pre-merging
phase\cite{Lindblom92}. For this purpose it is important to complete
theoretical templates of gravitational waves in the pre-merging phase
as well as in the inspiraling phase.

Recently Bonazzola, Gourgoulhon and Marck have numerically calculated
quasi-equilibrium configurations of irrotational BNSs in general
relativity\cite{BGM99}\footnote{The irrotational state is considered
to be a realistic state for binary neutron stars before
merger\cite{Kochanek92,BC92}.\\
$\ddagger$Present address: D\'epartement d'Astrophysique
	Relativiste et de Cosmologie, UPR 8629 du C.N.R.S.,
	Obserbatoire de Paris, F-92195 Meudon Cedex, France}.
Although their results seem to be
reasonable, we do not have any analytic solutions in general
relativity in order to check the validity of their results.

On the other hand, Ury\=u and Eriguchi have numerically constructed
stationary structures of irrotational BNSs in Newtonian
gravity\cite{UE98}. We can use semi-analytic solutions produced by
Lai, Rasio and Shapiro (hereafter LRS)\cite{LRS94} for checking the
validity of the results.

However, in numerical solutions of Ury\=u and Eriguchi, the velocity
component along the orbital axis exists while in those of LRS such a
component is assumed to be zero from the beginning. When we extend the
analytic solutions to the general relativistic ones\cite{Lom97}, we
should include this velocity component. This is because in the
numerical calculation, there is a possibility to obtain another
solution although the binding energy of a BNS is almost the same
value, and to lead a different conclusion\cite{MMW99}.

In order to include the velocity component along the orbital axis, we
solve the equation of continuity with the other basic equations. The
method we use in this Letter is that we seek a solution to an
equilibrium of irrotational binary polytropic stars in Newtonian
gravity by expanding all physical quantities in a power of $\e \equiv
a_0/R$, where $R$ and $a_0$ are the separation of the binary system
and the radius of each star for $R=\infty$. We extend the method
developed by Chandrasekhar more than 65 years ago for corotating
fluids\cite{Ch33} to the one for irrotational fluids.

Although a binary system consists of two stars, we pay particular
attention to one of two stars. We call it star 1 whose mass is $M_1$
and the companion one star 2 whose mass is $M_2$. In this Letter, we
adopt two corotating coordinate systems. First one is $\bX$ whose
origin is located at the center of mass of the binary system. For
calculational convenience, we choose the orbital axis as $X_3$, and we
take the direction of $X_1$ from the center of mass of star 2 to that
of star 1. The second coordinate system is the spherical one $\br
=(r,\th,\vp)$ whose origin is located at the center of mass of star
1. We use units of $G=1$.


Since we treat irrotational fluids in Newtonian gravity, the basic
equations are the equation of state, the Euler equation, the equation
of continuity and the Poisson equation:
\beqa
  &&P=K \r^{1 +{1 \over n}}, \\
  &&\n \biggl[ \int {dP \over \r} -U +{1 \over 2} v^2 -\bv \cdot
  ({\bf \O} \times \br) \biggr] =0, \label{Eq:Euler} \\
  &&\n \cdot \bv =-\bigl( \bv -{\bf \O} \times \br \bigr)
  \cdot {\n \r \over \r}, \label{Eq:continuity} \\
  &&\D U =-4 \pi \r, \label{Eq:Poisson}
\eeqa
where $P$, $\r$, $n$, $U$ and $\O$ are the pressure, the density, the
polytropic index, the gravitational potential and the orbital angular
velocity, respectively. The gravitational potential $U$ is separated
into two parts, i.e., the contribution from star 1 to itself $U^{1 \ra
1}$ and that from star 2 to star 1 $U^{2 \ra 1}$.
$U^{2 \ra 1}$ is written as
\beqa
  &&U^{2 \ra 1} ={M_2 \over R} \sum_{l=0}^{\infty} (-1)^l \Bigl(
	{\a \over R} \Bigr)^l \xi^l P_l (\sin \th \cos \vp)
	\nonumber \\
  &&\hspace{10pt}+{3 \bI_{11}' \over 2R^3} \Bigl[ 1 -3\Bigl(
	{\a \over R} \Bigr) \xi \sin \th \cos \vp +O(R^{-2}) \Bigr]
	+\cdots. \label{Eq:U21}
\eeqa
Here, $P_l$ and $\bI_{11}$ denote the Legendre function and the
reduced quadrupole moment, and the superscript $'$ means the term
concerned with star 2. $K$ is a constant related to entropy and $\bv$
represents the velocity field in the inertial frame. In the
irrotational fluid case, we can express $\bv$ as a gradient of a
scalar function $\Phi$, i.e., $\bv =\n \Phi$.

Following the {\it Lane-Emden equation} we first express the density
as $\r =\r_c \Th^n$ with $\r_c$ being the central density.
By using $\a \equiv [ K(1+n) \r_c^{1/n -1}/(4\pi) ]^{1/2}$,
we also introduce $\xi=r/\a$.
We expand $\Th$ in a power series of a parameter $\e$ as $\Th
=\sum_{i=0}^{\infty} \e^i \Th_i$. Since the shape of star 1 is
spherical when $R$ is large, the lowest order term of $\Th$ is the
solution of the {\it Lane-Emden equation}. Then we expand $\Th_i$ by
spherical harmonics as $\Th_i =\sum_{l,m} {}^{(i)}\! \psi_{lm} (\xi)
Y_l^m (\th,\vp)$. The radius of a spherical star $a_0$ is given by
$a_0=\a \xi_1$ as usual. Note here that $\xi_1=\pi$ for $n=1$ case.

Now we consider the orbital motion of star 1. In the spherical
coordinate system, it becomes
\beq
  {\bf \O} \times \br =\O R \bigl( \t{\bf \O} \times \bxi
  \bigr)_{orb} +\O \a \bigl( \t{\bf \O} \times \bxi \bigr)_{fig},
  \label{Eq:orbit}
\eeq
where
\beqa
  \bigl( \t{\bf \O} \times \bxi \bigr)_{orb}&=&{1 \over 1+p}
  (\sin \th \sin \vp, ~\cos \th \sin \vp, ~\cos \vp), \\
  \bigl( \t{\bf \O} \times \bxi \bigr)_{fig} &=&(0, ~0, ~\xi \sin \th),
\eeqa
and $p\equiv M_1/M_2$. The first term on the right-hand side of Eq.
(\ref{Eq:orbit}) comes from the orbital motion of the center of mass
of star 1 and the second term comes from the fluid motion around the
center of mass of star 1.

Next, we rewrite the equation of continuity (\ref{Eq:continuity}) as
\beq
  \D \Phi =-n (\n \Phi -{\bf \O} \times \br) \cdot
	{\n \Th \over \Th}. \label{Eq:Phi}
\eeq
The condition for $\Phi$ at the stellar surface is
$(\n \Phi -{\bf \O} \times \br) \cdot (\n \Th) \bigl|_{surf} =0$,
since $\Th=0$ at the surface. We expand $\Phi$ also as $\Phi
=\sum_{i=0}^{\infty} \e^i \Phi_i$. The gradient of the lowest order
term of $\Phi$ should agree with the orbital motion of the center of
mass of star 1, i.e., $\O R (\t{\bf \O} \times \bxi)_{orb}$ because
when $R$ is large, the shape of star 1 is spherical and star 1 has
only the orbital motion of the center of mass in the inertial frame
with no intrinsic spin. This leads us to normalize $\Phi$ as $\t{\Phi}
=\e \Phi/(\O \a a_0)$. We again expand $\t{\Phi}_i$ by spherical
harmonics as $\t{\Phi}_i =\sum_{l,m} {}^{(i)}\! \phi_{lm} (\xi)
\t{Y}_l^m (\th, \vp)$.

The orbital angular velocity is derived from the
first tensor virial relation defined by\cite{LRS93}
\beq
  \int d^3 x {\p P \over \p x_1} =0, \label{Eq:firstTV}
\eeq
where $x_1 =r\sin \th \cos \vp$.
If we substitute Eq. (\ref{Eq:orbit}),
$\Th_0$ and $\Phi_0$ into Eq. (\ref{Eq:firstTV}), we obtain the
orbital angular velocity in the lowest order as $\O_0^2 =M_{tot}/R^3$
where $M_{tot}=M_1 +M_2$. Note here that we also expand $\O^2$  as
$\O^2= \sum_{i=0}^{\infty} \e^i \O_i^2$.

Finally, we express the gravitational potential by rewriting
   Eq. (\ref{Eq:Euler}) as
\beq
  U =K(1+n) \r^{1 \over n} +{1 \over 2} v^2
  -\bv \cdot ({\bf \O} \times \br) +U_0, \label{Eq:U}
\eeq
where $U_0$ is constant.
Substituting Eq. (\ref{Eq:U}) into the Poisson equation
(\ref{Eq:Poisson}), we obtain the equation to  determine 
the equilibrium figure as 
\beqa
  \a^2 \D \Th &=&-\Th^n -{1 \over 8\pi \r_c} \D \Bigl[ (\n \Phi)^2 -2
  (\n \Phi) \cdot ({\bf \O} \times \br) \Bigr].
  \label{Eq:Theta}
\eeqa

Now a solution can be obtained iteratively. Firstly, $\Th_i$ is
determined by demanding that the gravitational potential and its
normal derivative are continuous at the stellar surface\cite{Ch33},
that is, $U_{int}|_{\xi=\Xi} =U_{ext}|_{\xi=\Xi}$ and $\p U_{int} /\p
\xi|_{\xi=\Xi} =\p U_{ext} /\p \xi|_{\xi=\Xi}$, where $\Xi(\th,\vp)$
expresses the surface ($\Th (\xi=\Xi(\th,\vp)) =0$). Substituting
$\Th_i$ and Eq. (\ref{Eq:orbit}) into Eqs. (\ref{Eq:Phi}) and
(\ref{Eq:firstTV}), we obtain $\Phi_i$ and $\O_i^2$. After that we
substitute these equations into Eq. (\ref{Eq:Theta}) and derive
$\Th_{i+1}$. We continued this procedure up to order $\epsilon^6$
 in this Letter.


Since we would like to present almost {\it analytic} results, we only
calculate $n=1$ case in this Letter. For other polytropic indices, we
must solve differential equations {\it numerically}. The results of
these cases will be given in the subsequent paper\cite{TN99}.  The
density profile up to $O(\e^6)$ becomes
$\Th =\Th_0 +\e^3 \Th_3 +\e^4 \Th_4 +\e^5 \Th_5 +\e^6 \Th_6$,
where
\beqa
  \Th_0 &=&{\sin \xi \over \xi}, \\
  \Th_3 &=&{5 \over p} j_2 (\xi) P_2 (\sin \th \cos \vp), \\
  \Th_4 &=&-{7\xi_1 \over 3p} j_3 (\xi) P_3 (\sin \th \cos \vp), \\
  \Th_5 &=&{9\xi_1^2 \over (15-\xi_1^2)p} j_4 (\xi)
	P_4 (\sin \th \cos \vp), \\
  \Th_6 &=&{225 \over 7p^2 \xi_1^2} j_2 (\xi)
	P_2 (\sin \th \cos \vp)
  +\six \psi_{22} (\xi) P_2^2 (\cos \th) \cos 2\vp \nonumber \\
  &&+{405 \over 7(15-\xi_1^2) p^2} j_4 (\xi)
	P_4 (\sin \th \cos \vp) \nonumber \\
  &&-{11\xi_1^3 \over (105-10\xi_1^2)p} j_5 (\xi)
	P_5 (\sin \th \cos \vp).
\eeqa
Here $j_l$ and $P_l^m$ denote the spherical Bessel function and the
associated Legendre function. The function $\six \psi_{22}$ is
obtained by solving a differential equation:
\beqa
  &&\Bigl[ {1 \over \xi^2} {d \over d\xi} \Bigl( \xi^2 {d \over d\xi}
	\Bigr) -{6 \over \xi^2} +1 \Bigr] \six \psi_{22} (\xi)
	\nonumber \\
  &&={2(1+p) \over p\xi_1} \Bigl[ {1 \over \xi^2} {d \over d\xi} \Bigl(
	\xi^2 {d \over d\xi} \Bigr) -{6 \over \xi^2} \Bigr]
	\four \phi_2 (\xi),
\eeqa
where $\four \phi_2 (\xi)$ will be defined later.
We point out that the terms $\Th_1$ and $\Th_2$ disappear in
the equation for $\Th$
since there are no
tidal terms to produce $\Th_1$ and $\Th_2$ in the gravitational
potential.

We can write the velocity potential as
$\t{\Phi} =\t{\Phi}_0 +\e^4 \t{\Phi}_4 +\e^5 \t{\Phi}_5 +\e^6 \t{\Phi}_6$,
where
\beqa
  \t{\Phi}_0 &=&{1 \over 1+p} \xi \sin \th \sin \vp, \\
  \t{\Phi}_4 &=&\four \phi_2 (\xi) P_2^2 (\cos \th) \sin 2\vp, \\
  \t{\Phi}_5 &=&\five \phi_3 (\xi) \Bigl[ P_3^1 (\cos \th) \sin \vp
  	-{1 \over 2} P_3^3 (\cos \th) \sin 3\vp \Bigr], \\
  \t{\Phi}_6 &=&\six \phi_4 (\xi) \Bigl[ P_4^2 (\cos \th) \sin 2\vp
	-{1 \over 4} P_4^4 (\cos \th) \sin 4\vp \Bigr].
\eeqa
The functions $\four \phi_2$, $\five \phi_3$ and $\six \phi_4$ are
determined by solving following differential equations:
\beqa
  &&\Bigl[ {d^2 \over d\xi^2} +\Bigl( {2 \over \xi}
	+{\Th_0' \over \Th_0} \Bigr) {d \over d\xi}
	-{6 \over \xi^2} \Bigr] \four \phi_2
	=-{5 j_2 (\xi) \over 2p \xi_1 \Th_0 (\xi)}, \\
  &&\Bigl[ {d^2 \over d\xi^2} +\Bigl( {2 \over \xi}
	+{\Th_0' \over \Th_0} \Bigr) {d \over d\xi}
	-{12 \over \xi^2} \Bigr] \five \phi_3
	=-{7 j_3 (\xi) \over 12p \Th_0 (\xi)}, \\
  &&\Bigl[ {d^2 \over d\xi^2} +\Bigl( {2 \over \xi}
	+{\Th_0' \over \Th_0} \Bigr) {d \over d\xi}
	-{20 \over \xi^2} \Bigr] \six \phi_4
  	={3\xi_1 j_4 (\xi) \over 4(15-\xi_1^2)p \Th_0 (\xi)},
	\nonumber \\
\eeqa
where $\Th_0'$ denotes $d\Th_0/d\xi$. Note that the terms
$\t{\Phi}_1$, $\t{\Phi}_2$ and $\t{\Phi}_3$ disappear in
the equation for $\t{\Phi}$. In Table I, we show the results of
the velocity potentials and their derivatives for an identical star
binary $(p=1)$. From the expressions of $\t{\Phi}_4$, $\t{\Phi}_5$ and
$\t{\Phi}_6$, it is clear that the velocity has non-zero components
along the orbital axis. Because since $\four \phi_2$ is not
proportional to $\xi^2$, there remains the velocity component $\p
\t{\Phi}_4 /\p \t{x}_3$ where $\t{x}_3=\xi \cos \th$\footnote{In the
case of $n=0$, $\four \phi_2$ is proportional to $\xi^2$.}. We can
also show the existence of the velocity components along the orbital
axis for $\t{\Phi}_5$ and $\t{\Phi}_6$. This analytic result does not
agree with the semi-analytic solution given by LRS\cite{LRS94} but
qualitatively agrees with the numerical solutions given by Ury\=u and
Eriguchi\cite{UE98}.

The orbital angular velocity is calculated as
\beq
  \O^2 ={M_{tot} \over a_0^3} \e^3 \Bigl[ 1 +{9\e^5 \over 2a_0^2}
	\Bigl( {\bar{\bI}_{11} \over M_1} +{\bar{\bI}_{11}' \over M_2}
	\Bigr) +O(\e^7) \Bigr], \label{Eq:omega}
\eeq
where $\bar{\bI}_{11} \equiv \bI_{11}/\e^3$. The effect of the
quadrupole moments in $\O^2$ is $O(\e^8)$. One can see from
Eq. (\ref{Eq:U21}) that the quadrupole term is $O(\e^5)$ higher than
the monopole term since $\bI_{11}$ is $O(\e^3)$. In the case of a
different mass binary, i.e., a neutron star-black hole binary,
we can express the orbital angular velocity as
\beq
  \O^2 ={M_{tot} \over a_0^3} \e^3 \Bigl( 1 +{9 \e^5 \over 2a_0^2}
  {\bar{\bI}_{11} \over M_1} \Bigr),
\eeq
where we assume that a black hole is a point source.

Now it is ready to calculate relevant physical quantities.
The mass of star 1 is calculated as
$M_1 =\int_1 d^3 x \r =4\pi \r_c \a^3 \xi_1 [1+45\e^6/(2p^2 \xi_1^2)]$.
Since we consider the sequence for given baryon mass,
the mass of star 1 should be the same one, which yields
\beq
  \r_c =\r_{c0} \Bigl( 1 -{45 \over 2p^2 \xi_1^2} \e^6 \Bigr),
  \label{Eq:density}
\eeq
where $\r_{c0}$ is the central density for $R=\infty$.  We note that
the central density of star 2 should be written by changing $p$ into
$1/p$ in Eq. (\ref{Eq:density}). The equation (\ref{Eq:density}) means
that the central density of star 1 up to order $\epsilon^6$ decreases
from that of the spherical star as the separation decreases.  Although
we did not assume ellipsoidal figures like in Ref. \cite{LRS94}, the
result is essentially the same as that derived by Lai\cite{Lai96},
since the calculation of the central density up to $O(\e^6)$ includes
only quadratic terms. In both calculations the central density of
irrotational binary system decreases as $R^{-6}$. This dependence is
different from that of corotating ones, i.e.,
$R^{-3}$\cite{Ch33}.

\begin{figure}[h,t]
  \vspace{-0.6cm}
\epsfxsize 7cm
\begin{center}
\leavevmode
\epsfbox{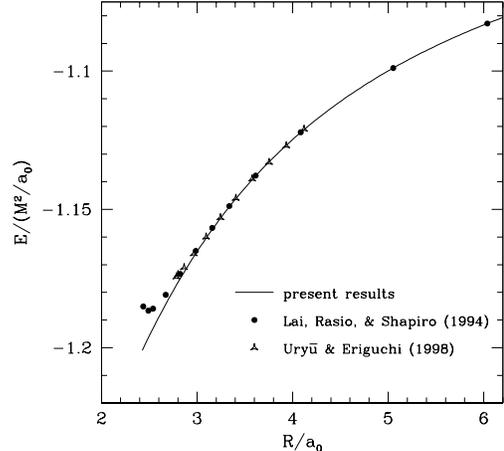}
  \vspace{-0.8cm}
\end{center}
\caption{The total energy as functions of the orbital separation for
an identical star binary $(M_1=M_2=M)$. Solid line denotes our
result. Filled circles and open triangles are the results of LRS (1994)
and Ury\=u and Eriguchi (1998).}
\label{fig1}
\end{figure}

Next, we calculate the total energy of the binary system, which is
written as $E =\Pi_{tot} +(W_{self})_{tot} +(W_{int})_{tot} +T_{tot}$,
where $\Pi_{tot}$, $(W_{self})_{tot}$, $(W_{int})_{tot}$ and $T_{tot}$
denote the total internal energy, the total self-gravity energy, the
total interaction energy and the total kinetic energy, respectively.
They are given as
\beqa
  \Pi_{tot} &=&{M_1 M_2 \over 4a_0} \Bigl( p+{1 \over p} \Bigr)
  \Bigl[ 1 -5 \Bigl( {15 \over \xi_1^2} -1 \Bigr) \e^6 \Bigr], \\
  (W_{self})_{tot} &=& {M_1 M_2 \over 4a_0} \Bigl( p +{1 \over p} \Bigr)
  \Bigl[ -3 +7 \Bigl( {15 \over \xi_1^2} -1 \Bigr) \e^6 \Bigr], \\
  (W_{int})_{tot} &=& -{M_1 M_2 \over a_0} \e -\e^6 {3M_1 M_2 \over
  2a_0^3} \Bigl( {\b{\bI}_{11} \over M_1} +{\b{\bI}_{11}' \over M_2}
  \Bigr), \\
  T_{tot} &=&{M_1 M_2 \over 2a_0} \e +\e^6 {9M_1 M_2 \over 4a_0^3}
  \Bigl( {\b{\bI}_{11} \over M_1} +{\b{\bI}_{11}' \over M_2} \Bigr),
\eeqa
where the reduced quadrupole moment of star 1 is calculated as
$\b{\bI}_{11} =(2M_1 a_0^2/3p)(15/\xi_1^2 -1)$.
The reduced quadrupole moment of star 2 is expressed by changing $M_1$
into $M_2$ and $p$ into $1/p$. We note that
the above energies  satisfy  the virial equation for $n=1$;~ i.e. 
 $3 \Pi_{tot} +(W_{self})_{tot} +(W_{int})_{tot} +2T_{tot} =0$.
Accordingly, the total energy is given as
\beqa
  &&E ={M_1 M_2 \over a_0} \Bigl[ -{1 \over 2} \Bigl( p +{1 \over p}
  \Bigr) -{\e \over 2}
  +\Bigl( {15 \over \xi_1^2} -1 \Bigr) \Bigl( p +{1 \over p} \Bigr) \e^6
  \Bigr].  \label{Eq:tot-ene} \nonumber \\
\eeqa
In the case of a different mass binary, we obtain the total energy
as
\beq
  E ={M_1^2 \over a_0} \Bigl[ -{1 \over 2} -{\e \over 2p} +{\e^6 \over p^2}
  \Bigl( {15 \over \xi_1^2} -1 \Bigr) \Bigr].
\eeq


\begin{figure}[h,t]
  \vspace{-0.6cm}
\epsfxsize 7cm
\begin{center}
\leavevmode
\epsfbox{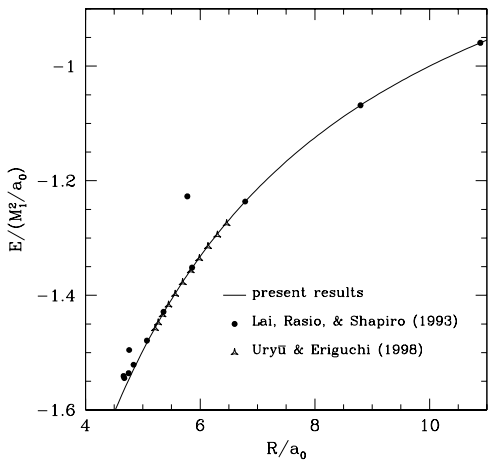}
  \vspace{-0.8cm}
\end{center}
\caption{The total energy as functions of the orbital
separation for a different mass binary $(p=0.1)$. Solid line
and open triangles are the same notations in Fig. 1.
Filled circles are the results of LRS (1993).}
\label{fig2}
\end{figure}

Our solution is correct if $\epsilon \ll 1$ so that it can be used to
check the validity of numerical solutions. For any numerical codes,
one can ask to solve an equilibrium for large $R$. One can compare
numerically derived density and velocity distribution with our almost
analytic solutions. However, for small $R$, our expansion up to $\e^6$
may not be enough. We can include the effects of the quadrupole terms
of stars in the physical values such as the total energy at order
$\e^6$. Since the next higher order terms are the octupole ones at
order $\e^8$ and their coefficients are order unity, we think that
this expansion converges for the effect of the deformation. However,
at order $\e^9$, there appears the spin kinetic energy term in the
total energy. There is a possibility to change the behavior of the
total energy for small $R$. Therefore, in order to apply our solution
in this case, further higher order calculations in our scheme as well
as the check of numerical codes using our almost analytic solutions
for large $R$ are urgently necessary.

In Figs. \ref{fig1} and \ref{fig2}, we show the total energy for
binary systems with $p=1$ and $p=0.1$ as functions of the orbital
separation for an identical star binary. We find from these figures
that the results of numerical and semi-analytic calculations coincide
with our analytic solutions up to $\e^6$ rather well in the region
$R/a_0>3$ for $p=1$ and $R/a_0>5$ for $p=0.1$ although they are quite
different in details.

We would like to thank K. Ioka and K. Nakao for useful discussions.
This work was partly supported by a Grant-in-Aid for Scientific
Research Fellowship (No.9402; KT) and Grant-in-Aid of Scientific
Research (No.09640351; TN) of the Japanese Ministry of Education,
Science, Sports and Culture.

\begin{table}
\caption{Velocity potentials and their derivatives for an identical star
binary $(p=1)$.
}
 \begin{center}
  \begin{tabular}{l|llllll}
  $\xi$&$\four \phi_2$&$\p_{\xi} \four \phi_2$&$\five \phi_3$&
  $\p_{\xi} \five \phi_3$&$\six \phi_4$&$\p_{\xi} \six \phi_4$ \\ \hline
  0.0&0.00&0.00&0.00&0.00&~0.00&~0.00 \\
  0.5&2.46(-2)&9.86(-2)&9.64(-4)&5.80(-3)&-3.40(-5)&-2.73(-4) \\
  1.0&9.91(-2)&0.200&7.80(-3)&2.36(-2)&-5.52(-4)&-2.23(-3) \\
  1.5&0.226&0.307&2.68(-2)&5.50(-2)&-2.87(-3)&-7.83(-3) \\
  2.0&0.408&0.424&6.55(-2)&0.103&-9.40(-3)&-1.96(-2) \\
  2.5&0.652&0.556&0.133&0.171&-2.41(-2)&-4.13(-2) \\
  3.0&0.967&0.710&0.241&0.269&-5.33(-2)&-7.87(-2) \\
  $\xi_1=\pi$&1.07&0.760&0.282&0.303&-6.55(-2)&-9.34(-2) \\
  \end{tabular}
 \end{center}
 \label{table1}
\end{table}%

\end{document}